\documentclass[a4paper, USenglish, cleveref]{lipics-v2021}
\usepackage{xspace}
\usepackage{import}
\usepackage[colorinlistoftodos,prependcaption,textsize=scriptsize]{todonotes}
\newcommand{\Nat}{\mathbb{N}}
\newcommand{\Real}{\mathbb{R}}
\newcommand{\fac}{\phi}

\newcommand{\Instances}{\mathcal{I}}
\newcommand{\bipmatching}{Maximum Bipartite Matching\xspace}

\DeclareMathOperator{\LCA}{LCA}
\usepackage{faktor}

\DeclareMathOperator*{\E}{\mathbb{E}}

\renewcommand{\leq}{\leqslant}
\renewcommand{\le}{\leqslant}
\renewcommand{\geq}{\geqslant}
\renewcommand{\ge}{\geqslant}

\newcommand{\trueparagraph}[1]{\smallskip\noindent\textbf{#1}\xspace}%

\newcommand{\mnew}[1]{#1} %

\title{\mnew{Submodularity Gaps for Selected Network Design and Matching Problems}}

\author{Martin Böhm}{University of Wrocław, Poland}{boehm@cs.uni.wroc.pl}{https://orcid.org/0000-0003-4796-7422}{}
\author{Jarosław Byrka}{University of Wrocław, Poland}{jby@cs.uni.wroc.pl}{https://orcid.org/0000-0002-3387-0913}{}
\author{Mateusz Lewandowski}{University of Wrocław, Poland}{mlewandowski@cs.uni.wroc.pl}{https://orcid.org/0000-0003-2912-099X}{}
\author{Jan Marcinkowski}{University of Wrocław, Poland}{jasiekmarc@cs.uni.wroc.pl}{https://orcid.org/0000-0002-6517-0014}{}

\authorrunning{M. Böhm, J. Byrka, M. Lewandowski and J. Marcinkowski} %
\Copyright{Martin Böhm, Jarosław Byrka, Mateusz Lewandowski and Jan Marcinkowski} %



\funding{NCN grant number 2020/39/B/ST6/01641.}%

\hideLIPIcs  %
\nolinenumbers

\EventEditors{John Q. Open and Joan R. Access}
\EventNoEds{2}
\EventLongTitle{42nd Conference on Very Important Topics (CVIT 2016)}
\EventShortTitle{CVIT 2016}
\EventAcronym{CVIT}
\EventYear{2016}
\EventDate{December 24--27, 2016}
\EventLocation{Little Whinging, United Kingdom}
\EventLogo{}
\SeriesVolume{42}
\ArticleNo{23}

\begin{document}

\maketitle

\begin{abstract}
Submodularity in combinatorial optimization has been a topic of many studies and various algorithmic techniques exploiting submodularity of a studied problem have been proposed. It is therefore natural to ask, in cases where the cost function of the studied problem is not submodular, whether it is possible to approximate this cost function with a proxy submodular function. %

We answer this question in the negative for two major problems in
metric optimization, namely Steiner Tree and Uncapacitated Facility
Location. We do so by proving super-constant lower bounds on the
submodularity gap for these problems, which are in contrast to the known constant factor cost sharing schemes known for them. 
Technically, our lower bounds build on strong lower
bounds for the online variants of these two
problems. Nevertheless, online lower bounds do not always imply 
submodularity lower bounds. We show that the problem \bipmatching
does not exhibit any submodularity gap, despite its online
variant being only $(1-1/e)$-competitive in the randomized setting.
  
\end{abstract}

\section{Introduction}
\label{sec:intro}

Submodularity is an important concept in the context of optimization over a space of subsets of a given ground set~\cite{fujishige2005submodular, badanidiyuru2014fast}. 

In this work, we consider several standard combinatorial optimization
problems: Steiner Tree (ST for short), Uncapacitated Facility Location
(UFL), and \bipmatching (MBM). They can be understood as providing a
certain service to a specified set of requests (terminals in ST,
clients in UFL, \mnew{assigning clients to vendors for MBM}) located in
points of a discrete metric space \mnew{or in vertices of a graph}. We
study the cost of the optimal solution of the respective problem as a
function of the set of requests. The specific question we focus on is:
Can this function be closely approximated by a submodular function of
the set of requests?

\subsection{Motivation}

Numerous result on submodular optimization are known, for and introduction to this field we refer to~\cite{fujishige2005submodular}. Specifically maximizing submodular functions can be done efficiently~\cite{badanidiyuru2014fast}. What is more important for us is that various problems become (more) tractable if we assume certain (cost) functions being submodular. This is perhaps best visible in the context of Cost Sharing Schemes which we discuss in more detail in Section~\ref{cost_sharing}.

To give a particular example of a recent result, in the context of dividing discrete goods trying to maximize Nash social welfare, it was shown that the assumption of individual valuation functions being submodular allows for obtaining constant factor approximation algorithms, see~\cite{garg2020approximating} and~\cite{garg2022approximating}.

In the context of settings where our goal is to serve requests in groups and the cost of serving a group of requests is given by a function, the assumption of the function being submodular can be exploited. In particular, such submodularity was used in the contest of 'service over time' problems studied by Bosman and Olver~\cite{bosman2020improved}, who gave $O(\log \log n)$-approximation algorithm for a setting called Submodular Routing Problem. More recently Abbasi et al.~\cite{abbasi2022loglog} gave $O(\log \log n)$-approximation algorithm
for Submodular Facility Location. In both these settings, under standard complexity theoretic assumptions, without the assumption on the specific cost functions are submodular, the approximation factors would need to be at least logarithmic. However, knowing that the function can be closely approximated via some other submodular function would suffice for an improvement.

\subsection{Formal statement}\label{subsec:formal-statement}
We first recall the formal definitions of Steiner Tree, Uncapacitated Facility Location, and Maximum Matching.

\trueparagraph{Steiner Tree.} In the Steiner Tree problem, we are given a metric space $X$, a root $r$ and a set of terminals $T \subseteq X$.
The objective is to find the minimum-length tree $S$ in the metric space such that $r \in V(S)$ and $\forall t \in T\colon t \in V(S)$.
The length of a tree $S$ is the sum of lengths of all edges in $S$.

\trueparagraph{Uncapacitated Facility Location.} In UFL, we are presented with a
metric space $X$ and a set of potential facilities $F \subseteq X$, where each
facility $\fac \in F$ costs a non-negative amount $c(\fac)$ to (permanently)
open. Alongside the facilities, there is also a multi-subset of clients $C$ of
size $n$, where each client $c \in C$ is an element of $X$. A feasible solution
for UFL is a set of open facilities $F' \subseteq F$ and an assignment $a\colon
C \to F'$ of all clients to open facilities. A cost of such a solution is then
$\sum_{\fac \in F'} c(\fac) + \sum_{i \in C} d(i, a(i))$, where $d(i,a(i))$ is
the distance in the metric space $X$ from the client to its assigned facility.
Naturally, the goal is to find a feasible solution of minimal cost.

\trueparagraph{Unweighted Maximum Matching in Bipartite Graphs.}
\mnew{We use a less common formalization of this well-known problem to
align more closely with our setting. We are presented with an
unweighted bipartite graph $G$ with two vertex partitions $U \uplus
V$, where $V$, $|V| = 2^{|U|} - 1$, contains a unique vertex adjacent
to each possible subset of $U$. We are then given a multi-subset $R$
of requested vertices from $V$ which are allowed to be matched to
their adjacent vertices in $U$. Each vertex from $U$ can be matched to
at most one request. The goal is to find the matching of $U \uplus R$
of maximum size.}

\trueparagraph{Dependence on the set of requests.} Given a combinatorial optimisation problem $P$ whose instances involve \mnew{a
distance graph} $X$ (e.g., $X$ being a metric space) and a subset of this \mnew{graph} $R \subset X$ referred to as the set
of requests, let $\Instances(P)$ be the set of all instances of $P$. Let
moreover $\Instances(P)^{-R} = \faktor{\Instances(P)}{\sim_R}$ denote the
equivalence classes of $\sim_R$ (instances are in the same equivalence class if
they differ only in the set of requests). For each such class $\Instances^{-r}
\in \Instances(P)^{-R}$ we define $c_{\Instances^{-r}}(R) : 2^X \rightarrow
\Real$ as a function that assigns to a set of requests $R$ the cost of the
optimal solution to problem $P$ on instance obtained from $\Instances^{-r}$ by
fixing the set of requests to $R$.

Note that for the Steiner Tree problem \mnew{and \bipmatching} the instance is composed of only a graph and a
set of \mnew{requests}, and hence it is natural to fix a single graph and study the
cost of the solution as a function of the set of \mnew{requests}. For the UFL problem,
an instance is composed of a metric, potential facilities, and clients. Here we
consider fixing both the metric and the set of potential facilities (with their
costs) and study the cost of the solution as a function of the set of clients.

A function $g\colon 2^X \to \Real$ is \emph{submodular} if it satisfies
\begin{equation}\label{eq:submod}
\forall A,B \subseteq X \colon \quad g(A) + g(B) \geq g(A \cup B) + g(A \cap B).
\end{equation}
\trueparagraph{Submodularity gap.} A function $f$ can be $\lambda-$approximated by a submodular function iff there exists a submodular function $g$ such that $f(S) \leq g(S) \leq \lambda \cdot f(S)~~\forall S \subseteq X$.
Let submodularity gap of $f$ be $SG(f)$ equal the minimal value $\lambda$ such that $f$ is $\lambda-$approximated by a submodular function. For an optimization problem $P$, we define its submodularity gap as
\[
SG(P) = \max \{SG(c_{\Instances^{-r}}) : \Instances^{-r} \in \Instances(P)^{-R}\}.
\]

\subsection{Relation with cost sharing schemes} \label{cost_sharing}

There is a natural relationship between submodularity of the cost function and the existence of cross-monotone cost sharing schemes studied by Pál and Tardos~\cite{pal2003group}. In the context of cost sharing one asks for both a solution to the studied optimisation problem and a distribution of the cost of the computed solution among the agents receiving the service.
Again, we may consider these cost allocations as a function of the set of requests (agents requesting service). We say the cost sharing scheme is cross-monotone if the cost assigned to a request is a monotone function of the set of requests (i.e. the larger the set the smaller the individual cost shares).

Submodularity of the cost function is very similar, it also encodes the ``economy-of-scale'' but on a global level: the increase of the total cost of the solution from adding one more request grows less if the request is added to a greater set. It is well known that submodularity of the cost function implies existence of an efficient cross-monotone cost sharing scheme, namely it suffices to consider adding requests one by one in a random order and allocate to each request the expected marginal contribution of this request. Moreover, it can be easily verified that if the submodularity gap of the objective function is at most $\lambda$, than the above described random order marginal contribution method yields cross-monotone distribution of a $1/\lambda$ fraction of the cost. Is the converse true? How much would we need to relax the approximation for a reduction in the other direction to hold?

For the problems studied in this paper the cost sharing situation is the following. For Uncapacitated Facility Location a 3-approximate cross-monotone cost sharing is possible~\cite{pal2003group}, and it is related to a primal-dual 3-approximation algorithm for UFL by Jain and Vazirani~\cite{jain2001approximation}. Better approximations for UFL are known, the current record is a 1.488-app\-roxi\-ma\-tion by Li~\cite{li20131}, but they cannot be simply turned into cross-sharing schemes. For the Steiner Tree problem there is a 2-approximate cross-monotone cost sharing scheme~\cite{jain2001applications} that can be seen as derived from a primal-dual 2-approximation algorithm of Goemans and Williamson~\cite{goemans1995general}. Again, a tighter 1.39-approximation is known for the Steiner Tree problem~\cite{byrka2013steiner}, but it cannot be turned into a cost-sharing scheme.

\subsection{Our results}

We show that, contrary to our initial intuitions, the submodularity gaps of \mnew{Facility Location and Steiner Tree} are not constant (as the relation with cost sharing schemes might suggest) but are logarithmic as they display a certain relation with the competitiveness of the online versions of the problems.

We first study the ``diamond'' instances from~\cite{imasewaxman} for the Steiner tree problem and obtain: 
\begin{theorem}
 The submodularity gap of Steiner tree is $SG(ST) = \Theta(\log n)$.
\end{theorem}
Using the fact that the cost of min cost Steiner tree is a 2-approximation of the cost of (Subset) TSP, we instantly obtain
\begin{corollary}
The submodularity gap of (Subset) TSP is $SG(TSP) = \Theta(\log n)$.
\end{corollary}

Then we consider the HST instances of UFL from~\cite{fotakis} to show
\begin{theorem}
 The submodularity gap of Uncapacitated Facility Location is $SG(UFL)= \Omega(\frac{\log n}{\log \log n})$.
\end{theorem}

The above results are obtained by considering the value of the cost function of the studied problem on carefully selected subsets of requests that display different levels of specificity.
A recursive argument is used to show that for a submodular overestimation of the cost function the value must grow with the increase of specificity despite the fact that the optimal cost of a solution for such a set remains low. This can be seen as related to amortised competitive analysis of online algorithms in which an algorithm accumulates budget for the future withdraw from a specific decision made in an early stage of the algorithm execution, however this relation is not being explicitly exploited in our bounds.

\mnew{Finally, we investigate the \bipmatching problem, which is
resolved as a classical online problem~\cite{GoelM08} but whose
generalizations (such as the AdWords problem) have immense practical
value and are not fully resolved yet~\cite{mehtasvv}.}

\mnew{For \bipmatching, we present evidence that the close relationship between the
submodularity gaps of all problems is not tied to the hardness of
their online variants. For MBM, the best deterministic competitive
ratio is $1/2$ and the best randomized competitive ratio is
$1-1/e$~\cite{GoelM08}; both results are tight~\cite{KarpVV}. In
contrast to this, the cardinality function of the set of requests $R$
is submodular, meaning:}

\begin{theorem}\label{thm:bipmatching}
The submodularity gap of \bipmatching is equal to $1$.
\end{theorem}

\newcommand\calP{\mathcal{P}}
\section{Submodularity Gap for Steiner Tree}\label{sec:st}

Our first result is a lower bound on the submodularity gap for Steiner
Tree, reusing the graph structure from the lower bound on the
competitive ratio of Online Steiner Tree \cite{imasewaxman}. As we
shall see, the graph structure is quite simple, in fact of treewidth
at most 2 (a series-parallel graph).

Our metric will arise from the \emph{diamond graphs} $D_k$. They have
an inductive definition: The diamond graph $D_0$ is equal to two
vertices $S$, $R$ and a single edge $e$, which we set to be of
\emph{level 0} and length $1$. The vertex $R$ will be the root for any
Steiner tree on this graph.

\begin{figure}
    \centering
    \includegraphics[]{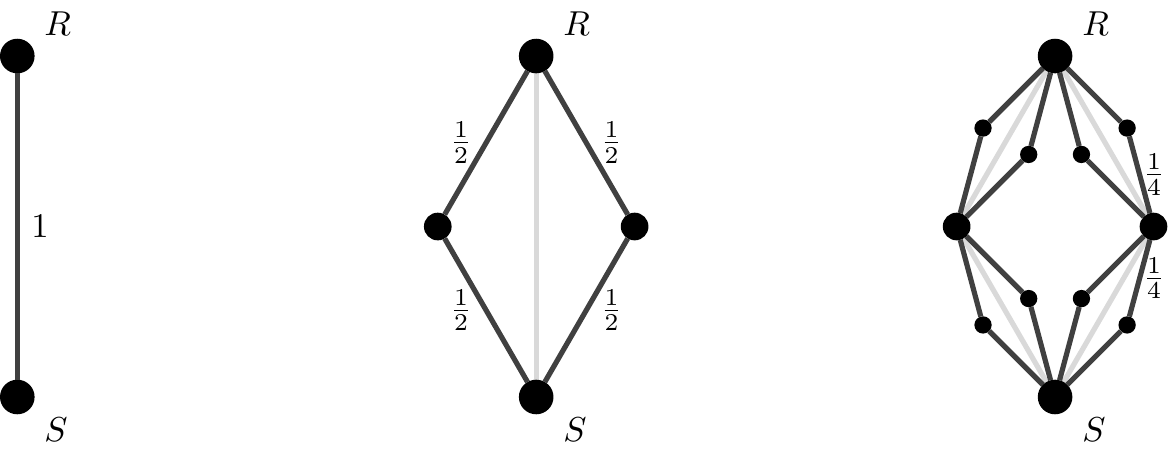}
    \caption{Diamond graphs $D_0$, $D_1$ and $D_2$. The graph $D_{k+1}$ is formed by replacing
    every edge in $D_k$ with a diamond with edge weights equal to $2^{-k}$. The
    distance between $S$ and $R$ is equal to $1$ in each
    graph.}\label{fig:steiner-diamonds}
\end{figure}

The graph $D_i$ can be derived from $D_{i-1}$ by first creating two
new copies of all edges of level $i-1$ (creating a multigraph in the
process), and then subdividing all the newly created edges,
which replaces them by two new edges (of level $i$ and half the length
of the previous edge) and a new vertex, which we also say is of
\emph{level $i$}. To distinguish the two newly created vertices
out of one edge of level $i-1$, we say one is a \emph{left} vertex and
the other is a \emph{right} vertex. See \Cref{fig:steiner-diamonds}
for an illustration containing $D_0$, $D_1$ and $D_2$.

Observe that due to our inductive definition, the graph $D_{j}$ is
contained inside the graph $D_i$ for $j \le i$; we will overload the
notation and refer to the set $D_j$ as the subset of vertices of $D_i$
with level at most $j$.

\trueparagraph{Notation.} In our arguments, we will often use
the name of a graph, such as a path $P = (V_P,E_P)$, interchangeably
with the vertex set $V_P$. This way we can talk about set operations
such as a union of two paths $P_1 \cup P_2$ in our submodularity
arguments.

A central object of the online lower bound of \cite{imasewaxman} are
$SR-$paths of the diamond graph. The length of any such path is equal
to $1$, which is equal to the optimal cost for the lower bound request
sequence, as the entire sequence lies on one specific $SR-$path. By
adversarially adding vertices of increasing levels into the online set
of terminals $K$, any online algorithm can be forced to pay $\Omega(\log
n)$ for its solution.

The lower bound on the submodularity gap will also focus on $SR-$paths
of a specific type, again defined inductively. The set $\calP_0$
contains only a single element, the unique $SR-$path in
$D_0$. To get $\calP_i$, we iterate over all $P \in \calP_{i-1}$
and extend $P$ into two paths $P_l, P_r$, which we insert into
$\calP_i$. The path $P_l$ is $P$ extended by all the left vertices of
level $i$ such that $P$ is still an $SR$-path, and $P_r$ is defined by
extending $P$ with all the right vertices of level $i$. We observe
that $|\calP_i|$ is exactly $2^i$ and that $\bigcup_{P \in \calP_i} P
= D_i$. Note also that each vertex of level $i$ belongs to only one
path in $\calP_i$. %

Due to the high symmetry of the diamond graphs, we can assume that for
a fixed $i$, every path $P \in \calP_i$ has the same value of $f(P)$.
Formally:

\begin{lemma}\label{lem:st:symmetry}
    Any submodular function $f$ that $\lambda$-approximates $c$ can be 
    transformed into a function $f^\text{sym}$ that also 
    $\lambda$-approximates $c$ and gives the same value to each $P \in
    \calP_i$.
\end{lemma}
\begin{proof}
    The diamond $D_i$ has $A(D_i) = 2{A(D_{i-1})}^4 = 
    2^{\frac{1}{12}(4^i-4)}$ automorphisms fixing $S$ and $R$. We define
    \[
        f^\text{sym}(L) = \frac{1}{|A(D_i)|} \sum_{\sigma \in A(D_i)} f(\sigma(L))
    \]
    for any set $L \subset V(D_i)$. The function $f^\text{sym}$ upper-bounds and
    $\lambda$-approximates $c$, since $c$ is invariant to the automorphisms.
    Moreover, $f^\text{sym}$ is a submodular function, because it is a sum of submodular 
    functions. All the paths in $P_i$ are in the same orbit of $A(D_i)$, so the function
    $f^\text{sym}$ assigns the same value to them.
\end{proof}

We will make use of \Cref{lem:st:symmetry} and use a variable
$f_i$ for the value of $f$ on any path $P_i$. As for the cost of the
optimal Steiner tree solution, it is easy to see that any path $P$
from $\calP_i$ has optimal cost $c(P) = 1$, as all vertices of $P$ lie
on a single path which contains both $S$ and $R$, making the cost
exactly $1$.

\subsection{Estimating the bound}\label{sec:st:computation}

Let $t_j$ be the optimum cost of $c(D_j)$, equal to the
cost of the spanning tree (a spanning tree is
the same as a Steiner tree with every vertex being a terminal). Such a spanning
tree has exactly $|V(D_j)|-1$ edges, each of length $2^{-j}$. Moreover
\[
    \begin{aligned}
        |V(D_{j+1})| &= |V(D_j)| + 2|E(D_j)|, \\
        |E(D_{j+1})| &= 4|E(D_j)|.
    \end{aligned}
\]
This gives us 
\[
    |V(D_j)| = \frac{2(4^j-1)}{3} + 2,
\]
and
\begin{equation}\label{eq:st-tn}
t_j = \frac{2}{3} \cdot 2^j + \frac{1}{3} \cdot 2^{-j}.
\end{equation}

The main inequality arising from submodularity is the following:

\begin{equation}\label{eq:recursion}
t_j \le f(D_j) \le f(D_{j-1}) + 2^j (f_j - f_{j-1}).
\end{equation}

Before adapting \Cref{eq:recursion} to get a lower bound on $f_d$ (the value of $f(P_d)$
for the paths of depth $d$), we discuss the correctness of this
inequality. We can see that $D_j$ (as a set of vertices)
satisfies $D_j = D_{j-1} \cup \bigcup_{P \in \calP_{j}} P$, which can be
imagined as adding the $2^j$ paths from $\calP_j$, each of which contains a unique vertex of level $i$,
into $D_{j-1}$, until we reach $D_j$.

Given a single path $P_{j-1} \in \calP_{j-1}$ and the set $L$ of vertices of level $j$ that
extend $P_{j-1}$ into a path in $\calP_{j}$ -- denoting this new path
$P_{j}$ -- we claim that the following holds:
\begin{equation}\label{eq:single-step}
f(D_{j-1} \cup L) \le f(D_{j-1}) + \bigl(f(P_j) - f(P_{j-1})\bigr).
\end{equation}

The correctness of this equation is an immediate consequence of the
definition of submodularity (\Cref{eq:submod}) if we set $A = D_{j-1}$
and $B = P_j$, using that $D_{j-1} \cap P_j = P_{j-1}$, as the new
vertices $L$ are the only elements of $P_j$ not already present in
$D_{j-1}$.

Generalizing \Cref{eq:single-step}, we have that
\begin{equation}\label{eq:single-step-generic}
f(X \cup L) \le f(X) + \bigl(f(P_j) - f(P_{j-1})\bigr)
\end{equation}
for any set $X$ and $L$ such that $L \cap X = \emptyset$, $P_{j-1} \subseteq X$ and $X
\cap P_j = P_{j-1}$. Thus, we can start with $X = D_{j-1}$ and add
each set $L$ one by one, maintaining
\Cref{eq:single-step-generic} until we reach $D_j$ after $|\calP_j| = 2^j$
steps, proving \Cref{eq:recursion}.

For the base case of \Cref{eq:recursion}, i.e. when $j=0$, we use the
simple equality $f(D_0) = f_0$. We can now expand the recurrence
in \Cref{eq:recursion} into its full form:

\begin{equation*}
  t_k \le f(D_k) \le f_0 + 2^1(f_1 - f_0) + 2^2(f_2 - f_1) + \cdots + 2^k(f_k - f_{k-1}),
\end{equation*}

and we express it as a lower bound on $f_k$ by rearranging:
\begin{equation}
f_k \ge \frac{t_k}{2^k} + \frac{f_{k-1}}{2} 
        + \frac{f_{k-2}}{2^2} + \cdots + \frac{f_0}{2^k}.
\end{equation}

We now compute a lower bound on any $f_k$ that satisfies this
inequality. To do so, we state a technical lemma that will apply
both to our construction for Steiner Tree as well as to \Cref{sec:fl}.

\begin{lemma}\label{lem:estimating}
    For any two sequences ${\{t_k\}}_{k\in\Nat}$, ${\{f_k\}}_{k\in\Nat}$ satisfying
    \begin{equation}
        f_k \geqslant \frac{t_k}{2^k} + \frac{f_{k-1}}{2} 
        + \frac{f_{k-2}}{2^2} + \cdots + \frac{f_0}{2^k} 
        \qquad \forall k \in \Nat,
        \label{eq:steiner-f-bound}
    \end{equation}
    it holds that
    \[
      f_k \cdot 2^k \geqslant (t_k - t_{k-1}) + 2^1(t_{k-1}-t_{k-2}) + 2^2(t_{k-2}-t_{k-3}) +\cdots 
        + 2^{k-1}(t_1-t_0)+2^k t_0.
    \]
\end{lemma}

\begin{proof}
    Let $g_l = 2^l f_l$. By rewriting~(\ref{eq:steiner-f-bound}) we have
    \begin{equation*}
        g_l \geqslant t_l + g_{l-1} + \cdots + g_0
        \qquad
        \forall l \in \Nat.
    \end{equation*}
    Now we will prove by induction that for every $k \in \Nat$,
    \[
        g_k \geq 2^0 (t_k - t_{k-1}) + 2^1 (t_{k-1} - t_{k-2}) + \cdots 
        + 2^{k-1} (t_1 - t_0) + 2^k t_0.
    \]
    The base case with $n=0$ is trivial as $g_0 = t_0 = 2^0 t_0$. In the general
    case we have
    \begin{align*}
        g_k &\geq t_k + g_{k-1} + g_{k-2} + \cdots + g_0 \\
        &\geq
        t_k + 
        [2^0 (t_{k-1} - t_{k-2}) + 2^1 (t_{k-2} - t_{k-3}) + \cdots + 
            2^{k-2} (t_1 - t_0) + 2^{k-1}t_0] +\\
        &\hspace{10em} [2^0 (t_{k-2} - t_{k-3}) + \cdots + 
            2^{k-3} (t_1 - t_0) + 2^{k-2}t_0] + \cdots + t_0 \\
        &=t_k + 2^0(t_{k-1} - t_{k-2}) + [2^1 + 2^0] (t_{k-2} - t_{k-3}) +
            [2^2 + 2^1 + 2^0] (t_{k-3} - t_{k-4}) \\
            &\quad+ \cdots + [2^{k-2}+2^{k-3}+\cdots+2^0] (t_1 - t_0)
            + [2^{k-1}+2^{k-2}+\cdots+2^0]t_0 \\
        &=t_k + [2^1 - 1](t_{k-1}-t_{k-2}) + [2^2 - 1](t_{k-2}-t_{k-3}) +
            \cdots + [2^{k-1}-1](t_1-t_0)+[2^k-1] t_0 \\
        &=t_k - (t_{k-1} - t_{k-2}) - (t_{k-2}-t_{k-3}) - \cdots - (t_1 - t_0) \\
          &\quad+ 2^1(t_{k-1}-t_{k-2}) + 2^2(t_{k-2}-t_{k-3}) +\cdots 
          + 2^{k-1}(t_1-t_0)+[2^k-1] t_0 \\
        &=(t_k - t_{k-1}) + 2^1(t_{k-1}-t_{k-2}) + 2^2(t_{k-2}-t_{k-3}) +\cdots 
        + 2^{k-1}(t_1-t_0)+2^k t_0.
    \end{align*}
  \end{proof}

  We now plug in the sequence $\{t_k\}$ from \Cref{eq:st-tn} to finish our computations.
    For $j=1,\dots,k$ we have $2^{k-j}(t_{j}-t_{j-1}) = \frac{1}{3}2^k\left(1 -
    2^{-2k}\right)$. Hence,
    \[
        f_k \cdot 2^k \geq \frac{k}{3}2^k - 
        \left(\frac{2^k}{3}\sum_{k=1}^{k}2^{-2k}\right) + 2^k t_0
        =
        2^k \left[\frac{k}{3} + \frac{1-4^{-k}}{9} + 1\right].
      \]

      The lower bound $f_k = \Omega(k)$ above is expressed in terms of the depth $k$
      of the diamond graph $D_k$, which has $n = 2(4^k-1)/3 + 2$ vertices. Thus, the submodularity gap
      for Steiner tree on instances with $n$ vertices is at least $\Omega(\log n)$.

\section[Realising Θ{(logn)} submodular approximation for Steiner Tree problem]{Realising $\Theta(\log n)$ submodular approximation for the Steiner Tree problem}
In this section we will see that the proved lower bound on the Submodularity Gap
for Steiner Tree is tight. In other words, for any metric $X$ we will construct
a function $f^X: 2^X\to \Real$ that is submodular and $\log |X|$-approximates
$c_X$---optimum values of Steiner Tree with a given set of terminals.

First we need to see that the function $c_X$ is itself submodular if the
underlying metric is a tree (or more formally, the metric being a metric
completion of a tree with distance associated to each edge).
\begin{lemma}
    For a tree $T$ and set $L\subset T$ of terminals, let $c_T(L)$ be the cost of
    the optimal Steiner tree. The function $c_T$ is submodular.
\end{lemma}
\begin{proof}
    Take any two sets $\emptyset \neq S_1 \subset S_2 \subset T$ and any vertex
    $v \not\in S_2$. We will compare $\Delta(S_1, v) = c_T(S_1 \cup \{v\}) -
    c_T(S_1)$ and $\Delta(S_2, v) = c_T(S_2 \cup \{v\}) - c_T(S_2)$.

    For a set $L\subset T$ and a vertex $v$, let $d(v, L)$ denote the distance
    from the vertex $v$ to the subtree spanning $L$. If $v$ lies on a path
    between some $L$-terminals, $d(v, L) = 0$. Clearly $d(v, S_1) \geq d(v,
    S_2)$. It is also easy to see, that $\Delta(L, v) = d(v, L)$ for any $L$ and
    $v$, which tells us, that the function $c$ is submodular (adding $v$ to a
    larger set has a smaller marginal cost).
\end{proof}
The second ingredient are tree embeddings of Fakcharoenphol, Rao and
Talwar~\cite{FakcharoenpholRT03}. They prove that
\begin{theorem}[FRT~\cite{FakcharoenpholRT03}]
    For any metric $X$ there is a family $\mathcal{T}$ of trees satisfying:
    \begin{align*}
        \forall T\in\mathcal{T},\quad &\forall i,j\in X\colon\quad T_{i, j} \geq X_{i, j},\\
        &\forall i,j\in X\colon \E_{T\sim\mathcal{T}}[T_{i, j}] \leq O(\log |X|) X_{i, j}.
    \end{align*}
\end{theorem}
For a metric $X$ we can define the function $f: 2^{X}\to\Real$,
\[
    f(L) = \E_{T \sim \mathcal{T}}[c_T(L)].
\]
The function $f$ is a sum of submodular functions ${\{c_T\}}_{T\in\mathcal{T}}$,
which makes it submodular. Moreover, for any $L\subset X$,
\[
    f(L) \leq O(\log |X|)c_X(L).
\]
It is a little bit harder to prove that $f \geq c_X$ --- we need to look at the
structure of trees in the family $\mathcal{T}$. Each such $T$ is a $2$-HST, which
means that for any two vertices $i,j\in X$,
\[
    X_{i, j} \leq 2^{l(\LCA_T(i, j))} \leq T_{i,j},
\]
where $l(v)$ is the level of a vertex $v$ in $T$ with leaves being at level $0$,
and $\LCA_T(u, v)$ is the \emph{lowest common ancestor} of $u$ and $v$ in
$T$. Take any set $L \subseteq X$ of terminals and a subtree $T[L]$ spanning it in
$T$. We will build a spanning subgraph of $L$ in $X$ using $T[L]$ as follows:
\begin{enumerate}
    \item Find a pair of vertices $i,j\in L$ with highest $l(\LCA_T(i, j))$;
    \item Connect $i,j$ in $X$ at a cost $X_{i,j}\leq 2^{l(\LCA_T(i, j))}$;
    \item Recursively build spanning subgraphs for two subtrees of $T[L]$ split
    at $\LCA_T(i, j)$.
\end{enumerate}
Throughout our procedure we consume every internal vertex of $T[L]$ once and the
cost of the connection added to $X_{i,j}$ was always lower than the cost removed
from $T[L]$. Hence, for every $T\in\mathcal{T}$ and every $L\subset X$,
\[
    c_X(L) \leq c_T(L).
\]

\section{Submodularity Gap for Uncapacitated Facility Location}\label{sec:fl}

In this section, we prove a lower bound of $\Omega(\log n/ \log \log
n)$ for the submodularity gap of Uncapacitated Facility Location with
$n$ clients. Our bound follows from a lower bound construction for
Online Uncapacitated Facility Location, which is due to Fotakis
\cite{fotakis}.

\begin{figure}
    \centering
    \includegraphics[scale=0.9]{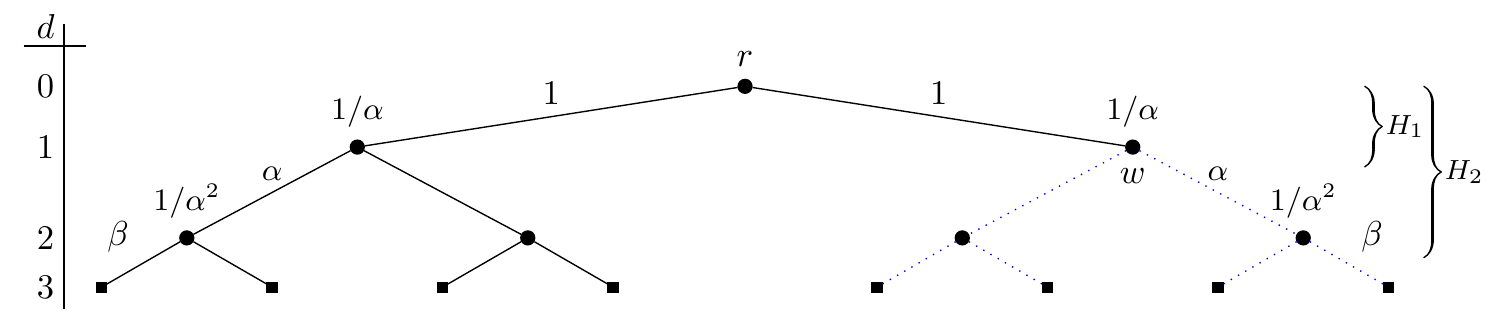}
    \caption{A 3-level HST $H_3$ with root $r$ and length parameter
      $\alpha$. The label of the edges indicates their length,
      whereas the label of the the vertices
      corresponds to the number of clients present in them.
      Facilities can only be opened in the leaves. We can also see
      the tree $H_1$ and $H_2$ as subtrees of $H_3$.
      The blue (dotted) edges correspond to the tree $H_3[w]$ for
      a vertex $w$. Note that the edges adjacent to the facilities
    have length $\beta = \alpha^2 + \frac{\alpha^3}{1-\alpha}$.}\label{fig:facloc}
\end{figure}

The lower bound of Fotakis \cite{fotakis} uses a \emph{hierarchically
well-separated tree} (HST) $H_d$ of depth $d$ as the graph inducing the
metric space $(V,d)$. A hierarchically well-separated tree is a full
binary tree with a root $r$ and a natural depth function for each
vertex and each edge in the graph (edges adjacent to the root are of
depth 0). As for the distances, every edge of depth $k$ has length
$\alpha^k$ for a parameter $0 < \alpha < 1$. See \Cref{fig:facloc}
for an illustration.

Note that again, a tree $H_d$ also contains all trees $H_j$ for $j \le
d$ as subtrees. Thus, we will again use $H_j$ to denote the set of
vertices inducing the tree $H_j$ within $H_d$. For one vertex of $v$,
we also employ the notation $H_d[v]$ to represent the subtree of all
descendants of the vertex $v$ within $H_d$.

So far, our definition of $H_d$ is standard in the literature.
However, for simplification of some calculations, we slightly increase
the length of all edges which are adjacent to the candidate
facilities, i.e. the leaves of the full tree $H_d$. More specifically,
we set the length of edges of depth $d-1$ to $\beta = \alpha^d + \frac{\alpha^{d+1}}{1 - \alpha}$
instead of the original $\alpha^d$. This way we guarantee that any
root-leaf path in the full tree $H_d$ is of length exactly $1/(1-\alpha)$.

The set $F$ of potential facilities is equal to the set of leaves of
$H_d$, each with the same opening cost $c(\fac)$, which we set to be
$\frac{2}{\alpha}$.

Having fixed our metric space to be the graph $H_d$, we consider the
metric invariant instances $\Instances(UFL)^R$ with at most $n$
clients. Our objective is to compute a lower bound on the value
$\max_{C} f(C)/c(C)$ for any valid choice of clients $C$.

For convenience, we will allow only one potential multiset for each
vertex $v$ in the HST $H$, and then speak of the cost $c(W)$, for a
set $W \subseteq V(H)$, equal to the cost of the instance where at
least one client appears on each vertex of $W$. However, unlike in the
case of the Steiner tree, there may be several clients on a single
vertex $v$. We restrict ourselves to instances where a vertex
of depth $k$ has exactly $1/\alpha^k$ clients on it, if it has
any. This way, the cost $c(W)$ of a set of vertices $W$ is always
well-defined.

Similarly to our approach to gaps for the Steiner tree, we wish to
evaluate the value of any submodular function $f$ on rooted paths as
well as on the subtrees $H_j$. We also use the notation of
\Cref{sec:st} and we use a graph variable such as $P$ to refer both to
a graph (a path) and to its vertex set.

Our set of rooted paths $\calP_i$ is defined to be the set of all
paths with exactly $i$ edges that start at the root vertex $r$. For
example, for depth $0$ we have $\calP_0 = \{H_0\} = \{\{r\}\}$.

\subsection{Estimating the bound}\label{sec:fl:calculation}

Overall, our approach to estimate the submodularity gap of Facility
Location is analogous to \Cref{sec:st:computation}, with necessary
changes due to a different graph and lower bound structure.

We first make an observation on bounding the connection cost of any
vertex $v$ of depth $k$ (with exactly $\frac{1}{\alpha^k}$ clients
on this vertex). {If all clients are connected to a facility in the
subtree $H_d[v]$, then the connection cost is exactly equal to
the number of clients times the distance to the nearest leaf, i.e.
$\frac{1}{\alpha^k} \cdot (\alpha^k + \alpha^{k+1} + \cdots + \beta) =
\frac{1}{1-\alpha}$.}

Next, let us first compute the optimum cost $c(H_j)$ of $H_j$ as well
as $c(P_j)$ for any path $P_j \in \calP_j$. For convenience, we denote
$t_j = c(H_j)$ and $o_j = c(P_j)$. For a single path $P_j$, the
optimal solution is to open any facility in the subtree below the path
$P_j$ and connect all clients on the path to it. The opening cost of
one facility is $2/\alpha$ and the connection cost {is again equal to}
$\frac{1}{1-\alpha}$ as discussed above, giving us
\begin{equation}\label{eq:on}
  o_j = \frac{2}{\alpha} + \frac{1}{1-\alpha}.
\end{equation}
For the optimum cost of $t_j = c(H_j)$, we open one facility for each
leaf vertex of $H_j$ (which is not a leaf of $H_d$ if $j < d$); more
specifically, for every leaf vertex $v$ of $H_j$, we open any one
facility in a leaf of $H_d[v]$. The number of leaves of $H_j$ is
exactly $2^j$, which sets the total opening cost to be
$\frac{2^{j+1}}{\alpha}$. For the connection costs, we can use our
observation above to express the connection cost as the number of
vertices in $H_j$ (which all have clients) times
$\frac{1}{1-\alpha}$. In total, we get
\begin{equation}
  t_j = \frac{2^{j+1}}{\alpha} + \frac{2^{j+1} - 1}{1-\alpha}.
\end{equation}

We briefly argue that our choices for opening and connections are
optimal. Since the facility opening cost is $2/\alpha$, it is never
advantageous to route a set of clients on a vertex $v$ of level $k$
outside the subtree $H_d[v]$, as routing them outside will cause the
total connection cost to rise at least above $\frac{1}{\alpha^d} \cdot
\alpha^{d-1} \cdot 2$, where we pay the connection cost twice as we
need to traverse two edges of length $\alpha^{d-1}$ to enter another
leaf that is not within $H_d[v]$. As
\[\frac{1}{\alpha^d} \cdot \alpha^{d-1} \cdot 2 = \frac{2}{\alpha}, \]
we can always open a new facility within the subtree of $H_d[v]$ and
route the client from $v$ to it.

We now show that this instance for Facility Location follows the same
formula as in \Cref{sec:st}. First, we take any path $P_j$ of length
$j$ and decompose it into a path $P_{j-1}$ of length $j-1$ (containing
the root) and a single vertex $v$ of depth $j$. Then, for this
decomposition, we have the following:
\begin{equation}\label{eq:fl-single-step}
f(H_{j-1} \cup \{v\}) \le f(H_{j-1}) + \bigl(f(P_j) - f(P_{j-1})\bigr).
\end{equation}

Again, this follows directly from the submodularity rule
\Cref{eq:submod}, if we set $A = H_{j-1}$ and $B = P_j$, using that
$H_{j-1} \cap P_j = P_{j-1}$.

\Cref{eq:fl-single-step} can again be generalized to yield
\begin{equation}\label{eq:fl-single-step-generic}
  f(X \cup \{v\}) \le f(X) + \bigl(f(P_j) - f(P_{j-1})\bigr)
\end{equation}

for any set of vertices $X$ such that $X \cap P_j = P_{j-1}$ and
$P_{j-1} \subseteq X$. Then, applying \Cref{eq:fl-single-step-generic}
$2^j$ times, once for each distinct vertex $v$ of level $j$, gives us
\begin{equation}
t_j \le f(H_j) \le f(H_{j-1}) + 2^j (f_j - f_{j-1}).
\end{equation}
We now apply \Cref{lem:estimating} for $f_j$ and $t_j$. To do so,
we first compute $t_j - t_{j-1}$:

\[t_j - t_{j-1} = \frac{2^{j+1}}{\alpha} + \frac{2^{j+1} - 1}{1-\alpha}
  - \left( \frac{2^{j}}{\alpha} + \frac{2^{j} - 1}{1-\alpha} \right) = 2^j \left(\frac{1}{\alpha} + \frac{1}{1-\alpha}\right).\]

Plugging this term into \Cref{lem:estimating} allows us to simplify:
    \begin{align*}
        f_d \cdot 2^d &\geq \sum_{j=1}^d 2^{d-j} (t_j - t_{j-1}) + 2^d t_0 \\
                      &\geq \sum_{j=1}^d 2^{d-j} 2^j \left(\frac{1}{\alpha} + \frac{1}{1-\alpha}\right) + 2^d\left( \frac{2}{\alpha} + \frac{1}{1-\alpha}\right) \\
                      &\geq 2^d \cdot d \cdot \left(\frac{1}{\alpha} + \frac{1}{1-\alpha}\right) + 2^d\left( \frac{2}{\alpha} + \frac{1}{1-\alpha}\right),
    \end{align*}
    ultimately simplifying to
    \begin{equation}
      f_d \ge \frac{d+2}{\alpha} + \frac{d+1}{1-\alpha}.
    \end{equation}
    To finish the computation, we need to evaluate the bound $f_d / o_d$, plugging in \Cref{eq:on}:
    \begin{equation}
      \frac{f_d}{o_d} \ge \frac{\frac{d+2}{\alpha} + \frac{d+1}{1-\alpha}}{\frac{2}{\alpha} + \frac{d+1}{1-\alpha}}.
    \end{equation}

If we create instances with $\alpha$ tending to zero, for
example by setting $\alpha = 1/\log n$, the first term of the fraction
will dominate the multiples of $\frac{1}{1-\alpha}$, giving us
$\lim_{\alpha \rightarrow 0^+} \frac{f_d}{o_d} = \frac{d+2}{2}$.

We measure the size of the instance by the number of clients $n$. The
maximum depth of the HST with $n$ clients is $d = O(\log n /
\log \log n)$, giving us a bound of $\Omega(\log n / \log \log n)$ on
the submodularity gap of any function $f$ for Uncapacitated Facility Location.

\section{Bipartite matching}

\mnew{We now shift our focus to \bipmatching, as formally defined
in~\Cref{subsec:formal-statement}, and show that its submodularity gap
is equal to $1$, meaning that the cardinality function of the matching
is submodular itself, with respect to the request set $R$.} Before we
are able to do that, let us prove a structural fact about maximum
matchings.

\begin{lemma}%
    \label{lem:matching-decomposition}
    In a bipartite graph $G = (U\uplus V, E)$ let $u\in U$ be a vertex which is
    matched in every maximum matching. Let $G'={(U \uplus V\cup \{v\}, E')}$ be
    the graph $G$ with added vertex $v$ and edges incident to it. The vertex $u$
    is matched by every maximum matching of $G'$.
\end{lemma}
\begin{proof}
    Let us fix a single maximum matching $M$. We can identify the set $X$ of
    vertices unmatched by $M$ and determine three sets: $\mathsf{EVEN}$
    (reachable by even-length alternating paths from $X$), $\mathsf{ODD}$
    (reachable by odd-length alternating paths from $X$) and $\mathsf{FREE}$
    (all the rest). In bipartite graphs no vertex is $\mathsf{ODD}$ and
    $\mathsf{EVEN}$ at the same time\footnote{In general graphs one needs to
    define $\mathsf{ODD}$ vertices as those reachable from $X$ by odd-length
    alternating paths but not reachable by even-length alternating paths. In
    bipartite graphs this exclusion is unnecessary, since such a vertex would
    belong to an odd-length cycle or an augmenting path.}.
    By~\cite[24.4b]{schrijver}, the set $\mathsf{EVEN}$ is exactly the set of
    vertices not matched by some maximum matching of $G$ (the set $D$ in
    Edmonds-Gallai decomposition). Consequently, $u \not\in \mathsf{EVEN}$. We
    will show that adding $v$ cannot move a vertex in $U$ from $\mathsf{ODD}
    \cup \mathsf{FREE}$ to $\mathsf{EVEN}$.

    After adding the vertex $v$ we can try to augment the matching $M$ to find
    the maximum matching of the larger graph. If there is no augmenting path,
    then $M$ is also a maximum matching of $G$. We look at vertices reachable
    from $v$ by alternating paths. In the decomposition for $G'$ these vertices
    become $\mathsf{ODD}$ (those that belong to $U$) and $\mathsf{EVEN}$ (those
    belonging to $V\cup \{v\}$).

    If there is an augmenting path $p = (v, \dots, w)$, we can apply it and
    obtain a new maximum matching $M'$. The vertex $w$ becomes matched but there
    are no newly unmatched vertices, so in the decomposition of $G'$ some
    vertices may transition from $\mathsf{ODD} \cup \mathsf{EVEN}$ to
    $\mathsf{FREE}$, \mnew{but they cannot switch their parity, as claimed.}
\end{proof}

\begin{lemma}%
    \label{lem:matching-submodular}
    For a bipartite graph $G = (U \uplus R, E_R)$, let $m_{U}{(R)}$ be the size
    of the maximum matching of $G$. The function $m_U$ is submodular.
\end{lemma}
\begin{proof}
    Let $R \subset S$. Adding one vertex $v \not\in S$ may increase the matching
    size by at most one. We will show that if $m_{R\cup {\{v\}}} = m_R$ then
    also $m_{S\cup {\{v\}}} = m_S$. This will give us $m_{S\cup {\{v\}}} - m_S
    \leq m_{R\cup {\{v\}}} - m_R$.

    $m_{R\cup {\{v\}}} = m_R$ means that in every maximum matching in $U \uplus
    R$ all the neigbours of $v$ are already matched. By repeatedly applying
    Lemma~\ref{lem:matching-decomposition}, all neigbours of $v$ are also
    matched in every maximum matching of $U \uplus S$, so $m_{S\cup\{v\}} =
    m_S$.
\end{proof}

\section{Concluding remarks}

\mnew{We have shown} logarithmic lower bounds on submodularity gap for both
Steiner Tree and Uncapacitated Facility Location. Note that our
construction shows that the gap for UFL manifests already for the case
when the underlying metric is a tree. On the other hand, Steiner Tree
on a tree metric has no submodularity gap, but the gap appears already
in series-parallel graphs (graphs of treewidth 2).

Our bounds build on lower bounds for competitive ratio of online
algorithms given requests adversarially. Note that online UFL becomes
easier (it is constant competitive) if the requests arrive in a random
order~\cite{meyerson2001online} and also in the incremental model
of~\cite{fotakis}, hence the submodularity gap concept appears to be
more related to the adversarial arrival model. We have used existing
online lower bounds to derive submodularity gap lower bounds. It
appears interesting to study whether one may derive arguments in the
opposite direction, e.g., new bounds on competitive ratio based on
bounds on the submodularity gap.

\mnew{Finally, we do concede that our lower bound results on the
submodularity gap of Steiner Tree and UFL are, in scientific
terminology, negative results and show that other approaches are
required to improve the bounds of~\cite{bosman2020improved}. 
Note however that these bounds are in contrast to the existence of 
cross-monotone cost sharing schemes and are therefore of independent interest.}
\mnew{As we have shown by our result on \bipmatching, the connection
between the submodularity gap of combinatorial optimization problems
and their online variants does not apply to all well-known problems,
and the notion of submodularity gap deserves to be investigated
further.}

\bibliography{main}

\end{document}